\documentclass[prd,aps,preprint,showpacs,nofootinbib,eqsecnum,superscriptaddress]{revtex4-1}

\usepackage{hyperref}
\usepackage{amsfonts,amssymb,amsmath,amsthm,amsxtra,mathtools,euscript,mathrsfs}

\usepackage{bm}
\usepackage{graphicx,color,xcolor,subfigure}
\usepackage{dcolumn}
\usepackage{bbm}
\usepackage{multirow}

\def\a{\alpha}
\def\b{\beta}
\def\g{\gamma}
\def\d{\delta}

\def\l{\lambda}
\def\r{\rho}
\def\s{\sigma}

\def\m{\mu}
\def\n{\nu}
\def\f{\phi}

\def\pa{\partial}

\def\mca{\mathcal{A}}
\def\mcb{\mathcal{B}}

\def\ti{\tilde}

\def\bs{\boldsymbol}

\begin{document}
\title{A Dark Energy model from Generalized Proca Theory}

\author{Chao-Qiang Geng}
\email[Electronic address: ]{geng@phys.nthu.edu.tw}
\affiliation{School of Fundamental Physics and Mathematical Sciences, Hangzhou Institute for Advanced Study, UCAS, Hangzhou 310024, China}
\affiliation{International Centre for Theoretical Physics Asia-Pacific, Beijing/Hangzhou, China}
\affiliation{Department of Physics, National Tsing Hua University, Hsinchu 300, Taiwan}
\affiliation{Synergetic Innovation Center for Quantum Effects and Applications (SICQEA), 
Hunan Normal University, Changsha 410081, China}
\author{Yan-Ting Hsu}
\email[Electronic address: ]{ythsu@gapp.nthu.edu.tw}
\affiliation{Department of Physics,
	National Tsing Hua University, Hsinchu 300, Taiwan}
\author{Jhih-Rong Lu}
\email[Electronic address: ]{jhih-ronglu@gapp.nthu.edu.tw}
\affiliation{Department of Physics, National Tsing Hua University, Hsinchu 300, Taiwan}
\author{Lu Yin}
\email[Electronic address: ]{yinlu@gapp.nthu.edu.tw}
\affiliation{Department of Physics, National Tsing Hua University, Hsinchu 300, Taiwan}
\begin{abstract}
We consider a specific  dark energy model, which only includes the Lagrangian up to the cubic order in terms of the vector field self-interactions in the generalized Proca theory.  We examine the cosmological parameters in the model  by using the data sets of CMB and CMB+HST, respectively. In particular, the Hubble constant is found to be $H_0=71.80^{+1.07}_{-0.72}$     ($72.48^{+0.72}_{-0.60}$) $\rm kms^{-1}Mpc^{-1}$ at $68\%$~C.L.  with CMB (CMB+HST), which would alleviate the Hubble constant tension. We also obtain that the reduced $\chi^2$ values in our model are close to unity when fitting with CMB and CMB+HST,  illustrating that our model is a good candidate to describe the cosmological evolutions
of the universe.

\end{abstract}
\maketitle
\section{Introduction}
Recent cosmological observations have shown that our universe is experiencing a late-time acceleration~\cite{Riess:1998cb,Perlmutter:1998np}. 
The simplest attempt to explain this phenomenon is to introduce the cosmological constant
$\Lambda$ as a repulsive force effectively, which  is also embedded in  the $\Lambda$CDM model~\cite{Amendola:2015,Weinberg:1972}. 
Although the $\Lambda$CDM model is one of the most successful cosmological model describing the large-scale structure of the universe, 
it fails to solve   the fine-tuning and 
coincidence problems, referred to as   the cosmological constant problem~\cite{Weinberg:1988cp,Peebles:2002gy,ArkaniHamed:2000tc}.

One alternative way to account for the accelerating universe is by introducing additional degrees of freedom in the gravitational theory~\cite{Copeland:2006wr}. This can be achieved by either modifying the geometric part  or involving some new fluids with negative pressure in the energy-momentum tensor
of the Einstein field equation. In particular, a wide class of dark energy models can be constructed by adding an additional scalar field $\f$, 
which contains a derivative coupling to the Ricci scalar $R$. The most general scalar-tensor theories with second-order equations of motion were derived by Horndeski in 1974~\cite{Horndeski:1974wa}, which  have been found to have numerous applications in cosmology, particularly in dark energy and inflation~\cite{Deffayet:2011gz,Charmousis:2011bf}.

If we now replace the scalar field by a massive vector field $A^{\m}$, the most general second-order field equations are called
generalized Proca theories~\cite{Heisenberg:2014rta,Jimenez:2016isa}. The application of this kind of theories to cosmology up to
the  sixth-order of  the vector field self-interactions 
in the Lagrangian has been studied in both background and perturbation levels, and  compared with the observational data~\cite{DeFelice:2016yws,DeFelice:2016uil,deFelice:2017paw,DeFelice:2020sdq,Heisenberg:2020xak}. It has been shown that there exists a de Sitter solution relevant to the late-time expansion in these theories. 
In addition, the authors in Ref.~\cite{DeFelice:2016yws} also proposed a  dark energy model, 
in which the solution always approaches a de Sitter fixed point.

Recently, there has been  the so-called Hubble tension in cosmology, indicating a mismatch between the local measurements and  early-time observations for the expansion rate of the Universe, i.e. the Hubble parameter $H_0$. The measurements from the Hubble Space Telescope (HST) has implied the value of  $H_0$ to be $74.03 \pm 1.42$ $\rm km\;s^{-1}Mpc^{-1}$~\cite{Riess:2019cxk}, whereas,  the CMB measurement together with the $\Lambda CDM$ model has given $67.4 \pm 0.5$ $\rm km\;s^{-1}Mpc^{-1}$~\cite{Aghanim:2018eyx}.
Many approaches in the literature have been presented to understand this tension, which can be divided into the early-time and late-time modifications
of general relativity. Examples for the former are the early dark energy scenario~\cite{Poulin:2018cxd,Karwal:2016vyq}, primodial magnetic fields \cite{Jedamzik:2020krr}, dark energy-dark matter interactions~\cite{Agrawal:2019dlm}, while the latter includes the modified gravity theory, which is our adopted approach.
In this study, we concentrate on one of the simplest dark energy model from the generalized Proca theory, in which we only consider up to the cubic order of the vector field self-interactions
 and present the numerical analysis of the model based on
Refs.~\cite{DeFelice:2020sdq}.

This paper is organized as follows. In Sec. \uppercase\expandafter{\romannumeral 2}, we present  our specific model based on 
the generalized Proca theories along with the  background equations of motion. 
In Sec. \uppercase\expandafter{\romannumeral 3}, we show the global fitting results. 
Our conclusion is given Sec. \uppercase\expandafter{\romannumeral 4}. At last, we demonstrate the tensor, vector and scalar perturbations for this model in the Appendix.


\section{Our model with background studies}

The action of the generalized Proca theory is given by~\cite{Jimenez:2016isa, Heisenberg:2014rta}
\begin{align}\label{GPaction}
S=\int d^{4}x \sqrt{-g} (\mathcal{L}+\mathcal{L}_{M}),
\end{align}
where $g$ is the determinant of the metric tensor $g_{\mu\nu}$,
$\mathcal{L}_M$ is the matter Lagrangian, and $\mathcal{L}$ is given by
\begin{align}\label{GPactionL}
\mathcal{L}=\sum^{6}_{i=2} \mathcal{L}_i,
\end{align}
with $ \mathcal{L}_i$ related to the vector field self-interactions, defined by
\begin{align}
\mathcal{L}_2&=G_{2}(X,F,Y),
\nonumber\\
\mathcal{L}_3&=G_3(X)\nabla_{\mu}A^{\mu},
\nonumber\\
\mathcal{L}_4&=G_4(X)R+G_{4,X}(X)[(\nabla_{\mu}A^{\mu})^2-\nabla_{\r}A_{\s}\nabla^{\s}A^{\r}],
\nonumber\\
\mathcal{L}_5&=G_5(X)G_{\m\n}\nabla^{\m}A^{\n}-\frac{1}{6}G_{5,X}(X)
[(\nabla_{\m}A^{\m})^3-3\nabla_{\m}A^{\m}\nabla_{\r}A_{\s}\nabla^{\s}A^{\r}+2\nabla_{\r}A_{\s}\nabla^{\g}A^{\r}\nabla^{\s}A_{\g}]
\nonumber\\
&-g_{5}(X)\tilde{F}^{\a\m}\ti{F}^{\b}{}_{\m}\nabla_{\a}A_{\b},
\nonumber\\
\mathcal{L}_6&=G_{6}(X)L^{\m\n\a\b}\nabla_{\m}A_{\n}\nabla_{\a}A_{\b}
+\frac{1}{2}G_{6,X}(X)\ti{F}^{\a\b}\ti{F}^{\m\n}\nabla_{\a}A_{\m}\nabla_{\b}A_{\n},
\end{align}
where 
 $X=-\frac{1}{2}A_{\mu}A^{\m}, F=-\frac{1}{4}F_{\mu\n}F^{\m\n}, Y=A^{\mu}A^{\n}F_{\m}{}^{\a}F_{\n\a}$, $G_{i,X}=\pa{G_i}/\pa X$, 
 $F_{\m\n}=\nabla_{\m}A_{\n}-\nabla_{\n}A_{\m}, \ti{F}^{\m\n}=\frac{1}{2}\epsilon^{\m\n\a\b}F_{\a\b}, and L^{\m\n\a\b}=\frac{1}{4}\epsilon^{\m\n\r\s}\epsilon^{\a\b\g\d}R_{\r\s\g\d}$, with $\nabla_{\m}$, $\epsilon^{\m\n\a\b}$, and $R_{\r\s\g\d}$ corresponding to the covariant derivative operator, Levi-Civita tensor and Riemann tensor, respectively.

\subsection{Background Equations of Motion}
For the homogeneity and isotropy of the universe, we consider the metric to be the flat Friedmann-Lemaitre-Robertson-Walker (FLRW) one,
given by
\begin{equation}\label{FLRW}
ds^2=-dt^2+a^{2}(t)\delta_{ij}dx^idx^j,
\end{equation}
with $a(t)$ the scale factor, and 
the Proca vector field $A^{\mu}$ 
\begin{align} \label{Afield}
A^{\m}=(\phi(t),0,0,0).
\end{align}
In Eq.~(\ref{GPactionL}), we only concentrate on the terms up to the cubic order with $G_4(X)=G_4$ being a constant.
As a result, 
the background equations of motion can be obtained by varying the action \eqref{GPaction}~\cite{DeFelice:2016yws,DeFelice:2016uil,deFelice:2017paw,DeFelice:2020sdq}, given by
\begin{align}
&G_2-G_{2.X}\f^2-3G_{3,X}H\f^3+6G_4H^2=\r_M,\label{eomg1}\\
&G_2-\dot{\f}\f^2G_{3,X}+2G_4(3H^2+2\dot{H})=-P_M, \label{eomg2}\\
&\f(G_{2,X}+3G_{3,X}H\f)=0.\label{eomA}
\end{align}
where $G_{i,X} = \pa G_i/\pa X$, and the dot denotes the derivative
with respect to cosmic time $t$.
%
Here, we note that  the perfect fluid 
has been taken into account. That is, the energy-momentum tensor for matter can be written as $T^{\m}{}_{\n}=\text{diag}(-\r_{M},P_{M},P_{M},P_{M})$
with $\r_{M}$ ($P_{M}$) representing the energy density (pressure).
The matter sector is assumed to be composed of non-relativistic matter ($m$) and radiation ($r$) with continuity equations read as:
	\begin{align}
	\dot{\r}_{m, r}+3 H(1+w_{m, r})\r_{m, r} =0  ,
	\end{align}
	where and the equations of state are defined by 
	\begin{align}
	w_{m, r}=\frac{P_{m, r}}{\rho_{m, r}}= 0, \frac{1}{3}.
	\end{align}
	The energy density and pressure then become $\r_{M}=\r_{m}+\r_{r}$ and $P_{M}=\rho_{r}/3$, respectively.

 For   dark energy to be dominated  in the late-time cosmological epoch, the amplitude of the temporal component of the vector field $\phi$ should increase as the Hubble parameter decreases. As suggested in Refs.~\cite{DeFelice:2016yws,DeFelice:2016uil,deFelice:2017paw}, we  can use the relation, given by 
\begin{align}
\f^p \propto H^{-1},
\end{align}
where $p$ is a positive constant. Consequently, the functions of $G_{2,3}$ can  be chosen to be the powers of $X$:
\begin{align}\label{G(X)}
G_2(X,F)=F+b_2 X^{p_2}, \quad G_3(X)=b_3X^{p_3}.
\end{align}
Besides, we let $G_4(X)=M_{pl}^2/2$, where $M_{pl}$ is the reduced Planck mass, in accordance with general relativity.
In order to satisfy \eqref{eomA}, there are some constraints on
the parameters of $b$ and $p$, given by
\begin{align}
p_3=\frac{1}{2}(p+2p_2-1),  
\label{bp1} \\
2^{p_3-p_2}p_2b_2+3p_3b_3(\f^p H)=0. \label{bp2}
\end{align}
In our calculation, we introduce a new free parameter $s$, defined by
	\begin{align}
	\label{s1}
	s\equiv\frac{p_2}{p}\,,
	\end{align} 
which is relevant to the background evolution and has been already fitted in the literature.
In particular, in Ref.~\cite{deFelice:2017paw} it is found that $s=0.254^{+0.118}_{-0.097}$ with the CMB data by Planck, while it
 shifted to  $s=0.16\pm{0.08}$ when the RSD data is included. 
For simplicity and illustrating our results, we fix the parameter of $s$ to be 0.25 by hand, and then
 choose a simple set:
\begin{align}\label{bpvalue}
p_2&=1, \quad p=4; \quad s=0.25\,.
\end{align}
Consequently, the resulting modified Friedmann equations become
\begin{align}
3M_{pl}^2H^2&=\r_M+\r_{DE},\\
M_{pl}^2(3H^2+2\dot{H})&=-P_M-P_{DE},
\end{align}
where $\r_{DE}$ ($P_{DE}$) is  the energy density  (pressure) of dark energy, given by
\begin{align}
\label{rde}
\r_{DE}&=-\frac{1}{2}b_2\f^2,\\
\label{pde}
P_{DE}&=\frac{1}{2}b_2\f^2+\frac{1}{3}b_2 \dot{\f}\f H^{-1}.
\end{align}
Note that
 $b_2$ and $b_3$ are related by
\begin{align}
b_3&=-\frac{4\sqrt{2}}{15(\f^4 H)}b_2
\end{align}
based on   \eqref{bp2}.

\subsection{Background Cosmological Evolutions}

To study the cosmological evolutions, it is convenient to introduce the density parameters, defined as
\begin{align}
\label{om}
\Omega_{i}=\frac{\rho_i}{3 M_{pl}^2 H^2} , 
\end{align}
where $i=m,r,DE$, and 
\begin{align}
\Omega_{m}+\Omega_{r}+\Omega_{DE}=1 .
\end{align}

	Taking derivatives of $\eqref{om}$ and using of \eqref{eomg2} and \eqref{eomA}, the equations of motion for the energy densities can be written as~\cite{DeFelice:2016yws}:
\begin{align}
\label{omdeprime}
\Omega^{\prime}_{DE}&=\frac{(1+s)\Omega_{DE}(3+\Omega_r-3\Omega_{DE})}{1+s\Omega_{DE}} ,\\
\label{omrprime}
\Omega^{\prime}_{r}&=-\frac{\Omega_r[1-\Omega_r+(3+4s)\Omega_{DE}]}{1+s\Omega_{DE}} ,
\end{align}
where a prime denotes the derivative  respect to the e-folding number
$N \equiv \ln a$. Using Eqs.~\eqref{om}-\eqref{omrprime}
and initial conditions of the density parameters, the evolutions of $\Omega_{m,r,DE}$ in \eqref{om} can be solved.
The equation of state for dark energy, which is defined by \eqref{rde} and \eqref{pde}, can also be written as:
\begin{align}
\label{wde}
w_{DE}\equiv \frac{P_{DE}}{\rho_{DE}}={-\frac{3(1+s)+s\Omega_r}{3(1+s\Omega_{DE})}}.
\end{align}

\begin{figure}[h]
	\centering
	\includegraphics[angle=-90,width=0.7 \linewidth]{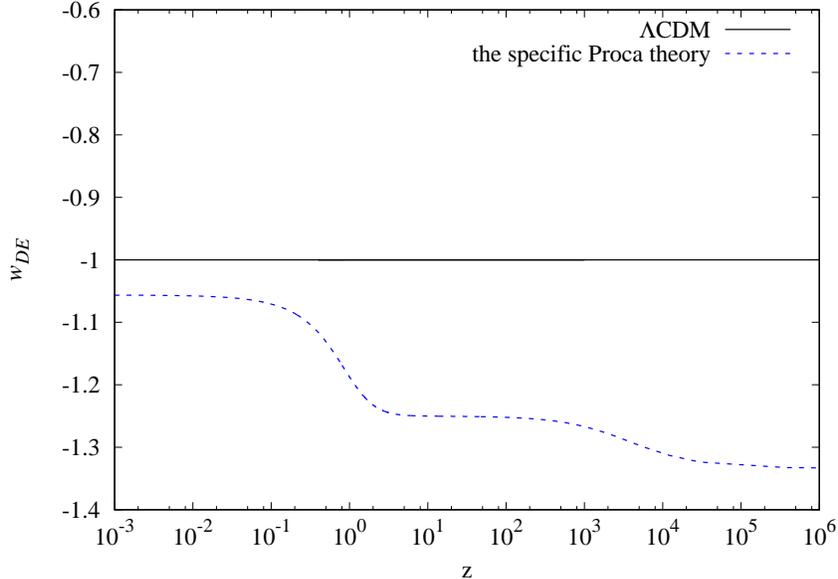}
	\caption{Evolutions of  equation of state for dark energy, $w_{DE}$, versus redshift for the model and $\Lambda$CDM}
	\label{fig:Wproca}
\end{figure}

\begin{figure}[h]
	\centering
	\includegraphics[angle=-90,width=0.7 \linewidth]{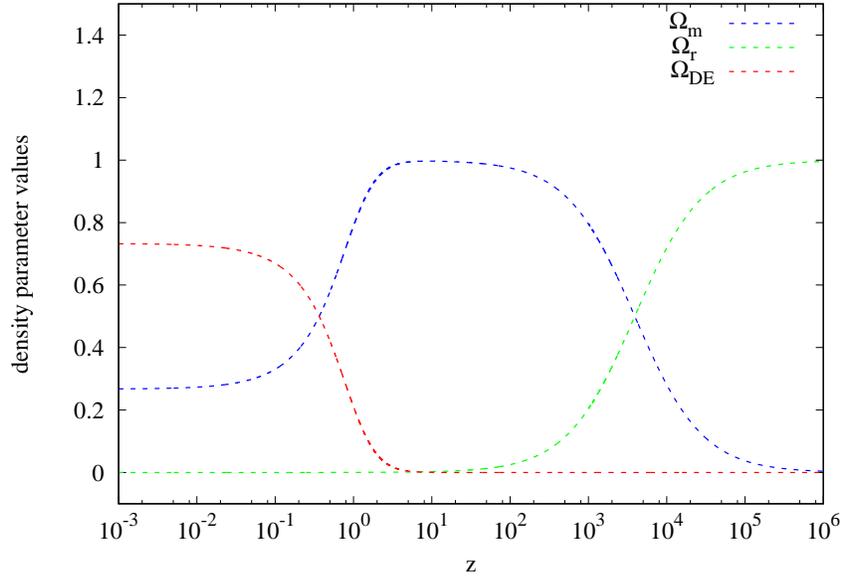}
	\caption{Evolutions of density parameters for matter ($\Omega_{m}$), radiation ($\Omega_{r}$) and dark energy ($\Omega_{DE}$) versus redshift for our model. The initial value are give by ($\Omega_{m}^0$, $\Omega_{r}^0$, $\Omega_{DE}^0$) = $(0.266, 6.83 \times 10^{-5}, 0.733)$, where ``0" denotes for the value at present.}
	\label{fig:Om}
\end{figure}

Using Eq.~(\ref{bpvalue}) 
in our model, Eqs.~\eqref{omdeprime}-\eqref{wde} become
\begin{align}
\label{omdeprime_s025}
\Omega^{\prime}_{DE}&=\frac{5\Omega_{DE}(3+\Omega_r-3\Omega_{DE})}{4+\Omega_{DE}} , \\
\label{omrprime_s025}
\Omega^{\prime}_{r}&=-\frac{4\Omega_r(1-\Omega_r+4\Omega_{DE})}{4+\Omega_{DE}} ,\\
\label{wde2}
w_{DE} &=-\frac{15+\Omega_r}{3(4+\Omega_{DE})},
\end{align}
respectively.
The evolutions of $w_{DE}$ and the density parameter versus the redshift $z$ for our specific model are plotted in Figs.~\ref{fig:Wproca} and \ref{fig:Om}, respectively. It can be seen that $w_{DE}$ evolves from $w_{DE}=-1-(4/3)s \approx-1.33$ in radiation era when $(\Omega_r,\Omega_{DE})=(1,0)$ to $w_{DE}=-1-s=-1.25$ in matter era when $(\Omega_r,\Omega_{DE})=(0,0)$, and $w_{DE}=-1.06$ at present, which shows an phantom-like behavior with $w_{DE}<-1$ ~\cite{{Heisenberg:2020xak}}.



\section{Global Fitting Results}

We use the {\bf CosmoMC}~\cite{Lewis:2002ah} and {\bf CAMB}~\cite{Lewis:1999bs} packages to study the constraints on cosmological parameters at the background level of our specific model in this section. The {\bf CosmoMC} package is a MCMC engine, which can be used to explore the parameter space based on maximum likelihood method.

To examine the behaviors of our model on the evolutions of the universe, we fit the models with the combination of 
the CMB and Hubble constant data sets. The CMB data include temperature 
and polarization angular power spectra from {\it Planck} 2018 with TT, TE, EE, low-$l$ polarization, 
and CMB lensing from SMICA~\cite{Aghanim:2018eyx,Aghanim:2018oex,Akrami:2019izv,Aghanim:2019ame}, while the Hubble constant of $H_0=74.03 \pm 1.42$ $\rm km\;s^{-1}Mpc^{-1}$ is from Hubble Space Telescope (HST) ~\cite{Riess:2019cxk}.
As we set $s=0.25$ and let the neutrino mass sum  be a free parameter, both our model and $\Lambda$CDM contain seven free parameters, where the priors are listed in Table~\ref{tab:prior}. 
To obtain the best fitted values of cosmological parameters, we use the  $\chi^2$ method
with 
\begin{eqnarray}
    \chi^2 = {\chi^2_{CMB}+ \chi^2_{HST}} \,,
\end{eqnarray}
Explicitly,  we take
\begin{eqnarray}
	\chi^2_{CMB} = (x^{th}_{i,CMB}-x^{obs}_{i,CMB})(C^{-1}_{CMB})_{ij}(x^{th}_{j,CMB}-x^{obs}_{j,CMB})\,,
\end{eqnarray}
where ``th'' and ``obs'' denote theory and observational values, respectively, $C^{-1}_{CMB}$ is the inverse covariance matrix, and $x_{i,CMB} \equiv (l_A(z_*),R(z_*),z_*)$ with the acoustic scale $l_A$ and the shift parameter $R$ at the photon decoupling epoch, $z_*$, defined by
\begin{eqnarray}
	l_{A}(z_*)&=&(1+z_*)\frac{\pi D_{A}(z_*)}{r_s(z_*)}\,,
	\nonumber\\
	R(z_*)&\equiv & (1+z_*)D_{A}(z_*)\sqrt{\Omega_m H^2_0}\,.
\label{DArs}
\end{eqnarray}
In Eq.~(\ref{DArs}),  $D_{A}(z)$ and  $r_s$ are the proper angular diameter distance and comoving sound horizon, given by
\begin{eqnarray}
	D_{A}(z)&=&\frac{1}{1+z}\int_{0}^{z} \frac{dz^{'}}{H(z^{'})}\,,
	\nonumber\\
	r_{s}(z)&=&\frac{1}{\sqrt{3}}\int_{0}^{1/(1+z)} \frac{da}{a^2 H(a) \sqrt{1 + (3\Omega^0_b/4\Omega^0_r)a}}\,,
\end{eqnarray}
respectively, where $\Omega^0_b$ and $\Omega^0_r$ present values of baryon and photon density parameters, respectively. 
For  $\chi^2_{HST}$ , we have
\begin{eqnarray}
\chi^2_{HST} = \frac{(74.03-H^{th}_0)^2}{1.42^2} \,,
\end{eqnarray}
where $H^{th}_0$ is the theoretical value of Hubble parameter in the model.

To compare the results between the models, we use the reduced  $\chi^2$, defined by
\begin{eqnarray}
	\chi^2_{reduced} = \frac{\chi^2}{\nu} \,,
\end{eqnarray}
where $\nu= N-n$ is the degrees of freedom, with ``N'' and ``n'' denote as the numbers of data points and free parameters, respectively.

\begin{table}[t]
	\begin{center}
	\caption{ Priors of the  cosmological parameters for our model and $\Lambda$CDM}
		\begin{tabular}{|c|c|} \hline
			Parameter & Prior
			\\ \hline
			Baryon density & $0.5 \leq 100\Omega_b^0 h^2 \leq 10$
			\\ \hline
			CDM density & $0.1 \leq 100\Omega_c^0 h^2 \leq 99$
			\\ \hline
			Optical depth & $0.01 \leq \tau \leq 0.8$
			\\ \hline
			Neutrino mass sum& $0 \leq \Sigma m_{\nu} \leq 2$~eV
			\\ \hline
			$\frac{\mathrm{Sound \ horizon}}{\mathrm{Angular \ diameter \ distance}}$  & $0.5 \leq 100 \theta_{MC} \leq 10$
			\\ \hline
			Scalar power spectrum amplitude & $2 \leq \ln \left( 10^{10} A_s \right) \leq 4$
			\\ \hline
			Spectral index & $0.8 \leq n_s \leq 1.2$
			\\ \hline
		\end{tabular}
		\label{tab:prior}
	\end{center}
\end{table}

\begin{figure}[t]
	\centering
	\includegraphics[width=1 \linewidth]{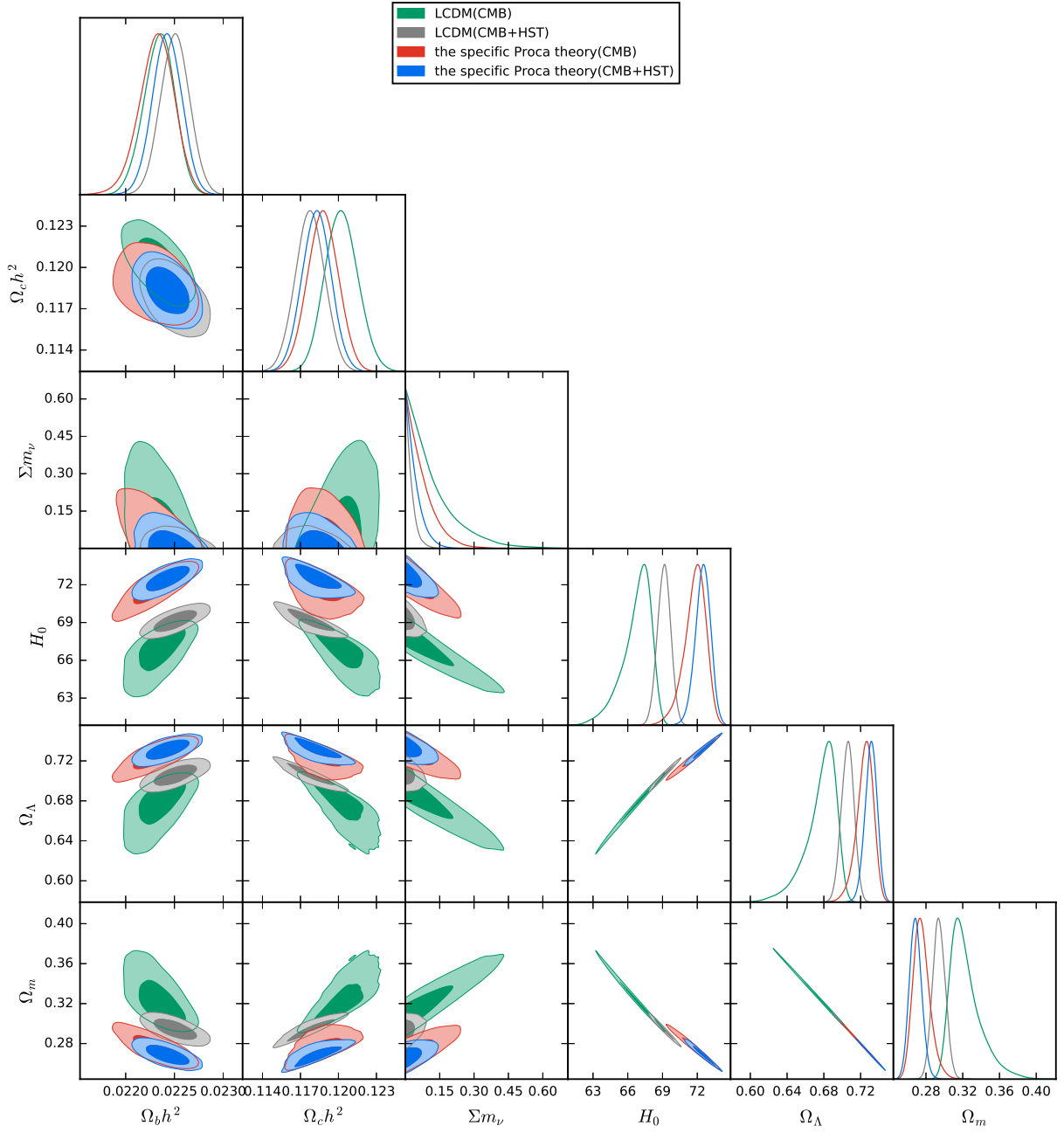}
	\caption{One and two-dimensional distributions of $\Omega_b^0 h^2$, $\Omega_c^0 h^2$, $\sum m_\nu$, $H_0$, $\Omega_{\Lambda}^0$, $\Omega_{m}^0$ for our model and $\Lambda$CDM  with the combined data of CMB and CMB + HST, where the contour lines represent 68$\%$~ and 95$\%$~ C.L., respectively.}
	\label{fig:fit}
\end{figure}

The constraints for the cosmological parameters of our model from the specific Proca theory 
with CMB and CMB+HST are plotted in  Fig.~\ref{fig:fit} and listed in  Table~\ref{tab:fit}.
It is given that $H_0=71.80^{+1.07}_{-0.72}$ ($72.48^{+0.72}_{-0.60}$) $\rm km\;s^{-1}Mpc^{-1}$ in our model and $H_0=66.75^{+1.52}_{-0.73}$ ($69.13\pm 0.57$) $\rm km\;s^{-1}Mpc^{-1}$ in $\Lambda$CDM when fitting with CMB (CMB+HST) at $68 \%$ C.L.  
It is interesting to see that
our model favors a larger $H_0$ even without including the HST data, while the addition of HST pulls $H_0$ to an even larger value, agreeing  better 
with the local measurements. 

While the early-time observation from Planck provides $H_0=67.4 \pm 0.5$ $\rm km\;s^{-1}Mpc^{-1}$ 
with the $\Lambda$CDM scenario~\cite{Aghanim:2018eyx}, late-time measurements of $H_0$ exceed early-time estimation to the extent more than $4 \sigma$~\cite{Riess:2020sih}. This Hubble constant tension may call for new physics beyond $\Lambda$CDM with different behavior in early and late times of the universe~\cite{Riess:2019cxk,Odintsov:2020qzd}. 
On the other hand, the generalized Proca theory, which has phantom-like behavior of $w_{DE}$ with $s>0$, naturally favors a  larger $H_0$ and is helpful to resolve  the $H_0$ tension~\cite{deFelice:2017paw,DeFelice:2020sdq,Heisenberg:2020xak}.
	
With the specific choice of the Lagrangian in the model, our result of  $H_0=71.80^{+1.07}_{-0.72}$  ($72.48^{+0.72}_{-0.60}$) with CMB (CMB+HST) matches the late-universe measurements of $74.03 \pm 1.42$ from HST~\cite{Riess:2019cxk}, $73.5 \pm 1.4$ from SH0ES~\cite{Reid:2019tiq}, 
$75.3^{+3.0}_{-2.9}$ from H0LiCOW~\cite{Wei:2020suh} and $73.9 \pm 3.0$ from Megamaser~\cite{Pesce:2020xfe}, in units of $\rm km\;s^{-1}Mpc^{-1}$
for $H_0$. This is caused by the phantom-like behavior in our model as plotted in Fig.~\ref{fig:Wproca}. 
 Moreover, the $H_0$ tension is reduced from $ 3.68\sigma$ ($ 3.20\sigma$) in $\Lambda$CDM to $1.30 \sigma$ ($0.99 \sigma$) in our model  with CMB (CMB+HST).

In addition, we obtain the best fits with $\chi^2_{Proca,reduced}=1.0960$ ($1.0959$) and $\chi^2_{\Lambda CDM,reduced}=1.0964$ ($1.1016$) when fitting with CMB (CMB+HST)\footnote{In a full analysis with $s$ left as a free parameter, this value might become a bit larger because the number of
free parameters is increased by 1.}, resulting in  that $| \chi^2_{Proca,reduced}-1|=0.0960$ ($0.0959$) and
$|\chi^2_{LCDM,reduced}-1|=0.0964$ ($0.1016$). As $\chi^2_{Proca,reduced}$  is closer to unity,  our model can well describe the late-time evolution of the universe.


\begin{table}[t]
	\newcommand{\tabincell}[2]{\begin{tabular}{@{}#1@{}}#2\end{tabular}}	
	\begin{center}
			\caption{Fitting results in our model and $\Lambda$CDM with the CMB and CMB + HST data sets,
			where the cosmological parameters are constrained at 68$\%$ C.L and the $H_{0}$ tension is compared with $H_0=74.03 \pm 1.42$ $\rm km\;s^{-1}Mpc^{-1}$ from HST, 
			by assuming the Gaussian distribution for 1D posterior.}
		\scalebox{0.9}{
			\begin{tabular} {|c|c|c|c|c|}
				\hline
				Parameter&
				\multicolumn{2}{c|}{
					CMB
				} 
				&
				\multicolumn{2}{c|}{
					CMB+HST
				} \\
				\hline
				Model&
				our model
				&
				$\Lambda$CDM
				&
				our model
				&
				$\Lambda$CDM
				\\
				\hline		
				{\boldmath$100\Omega_b^0 h^2   $} & 
				$2.232^{+0.019}_{-0.017}$& 
				$2.234^{+0.017}_{-0.014}$& 
				$2.242 \pm 0.015 $&
				$2.251 \pm 0.015 $
				\\
				
				{\boldmath$100\Omega_c^0 h^2   $} &
				$11.88 \pm 0.12 $&
				$12.03^{+0.12}_{-0.14}$&
				$11.83 \pm 0.11 $&
				$11.78 \pm 0.12 $
			    \\

				\boldmath$H_0 $            ($\rm kms^{-1}Mpc^{-1}$)        &
				$71.80^{+1.07}_{-0.72}   $& 
				$66.75^{+1.52}_{-0.73}   $&  
				$72.48^{+0.72}_{-0.60}   $& 
				$69.13\pm 0.57           $ 
				\\
				
				\boldmath$\Omega_{DE}^0     $& $0.7329^{+0.0103}_{-0.0073} $&
				$0.6766^{+0.0201}_{-0.0089} $&
				$0.7313^{+0.0071}_{-0.0062} $& 
				$0.7059^{+0.0068}_{-0.0070} $ 
				\\
				
				\boldmath$\Omega_{m}^0       $& $0.2754^{+0.0073}_{-0.0103}  $&
				$0.3234^{+0.0089}_{-0.0200}  $&
				$0.2687^{+0.0062}_{-0.0071}  $& 
				$0.2941^{+0.0070}_{-0.0068}  $ 
				\\
			    \hline
				\boldmath$H_{0}$ \bf{tension}& 
				1.30$\sigma$&
				3.68$\sigma$&
				0.99$\sigma$&
				3.20$\sigma$
				\\
				
				{\bf{Reduced} \boldmath $\chi^2_{best-fit}$ } & 
				$1.0960$& 
				$1.0964$&
				$1.0959$& 
				$1.1016$\\
				\hline
		\end{tabular}}
		\label{tab:fit}
	\end{center}
\end{table}

\begin{figure}[!htb]
	\centering
	\includegraphics[angle=-90,width=0.7 \linewidth]{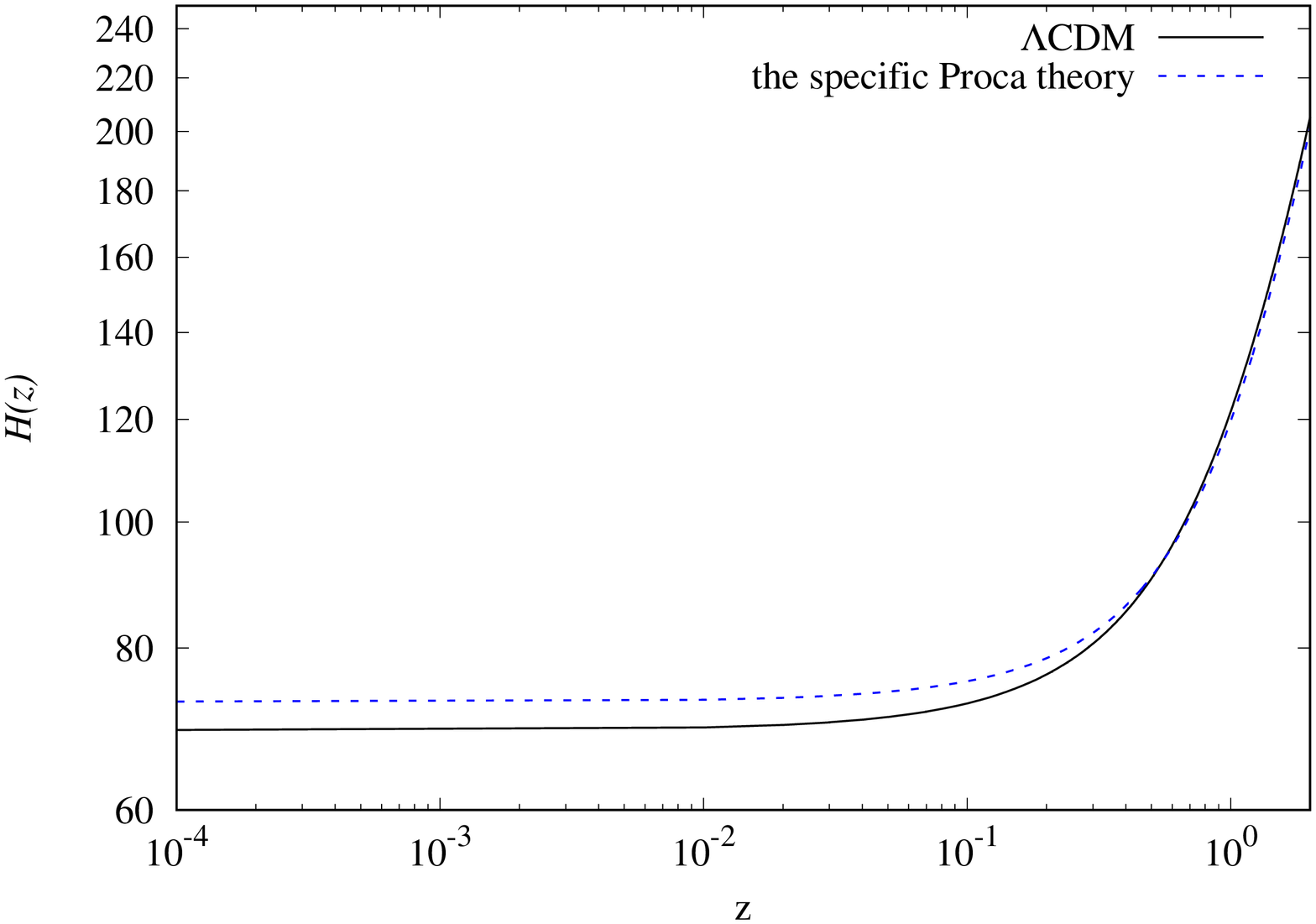}
	\caption{Evolution of Hubble parameter, $H(z)$, in the model and $\Lambda$CDM using the best-fit values from CMB+HST as initial conditions. }
	\label{fig:Hzproca}
\end{figure}

\begin{figure}[!htb]
	\centering
	\includegraphics[angle=-90,width=0.7 \linewidth]{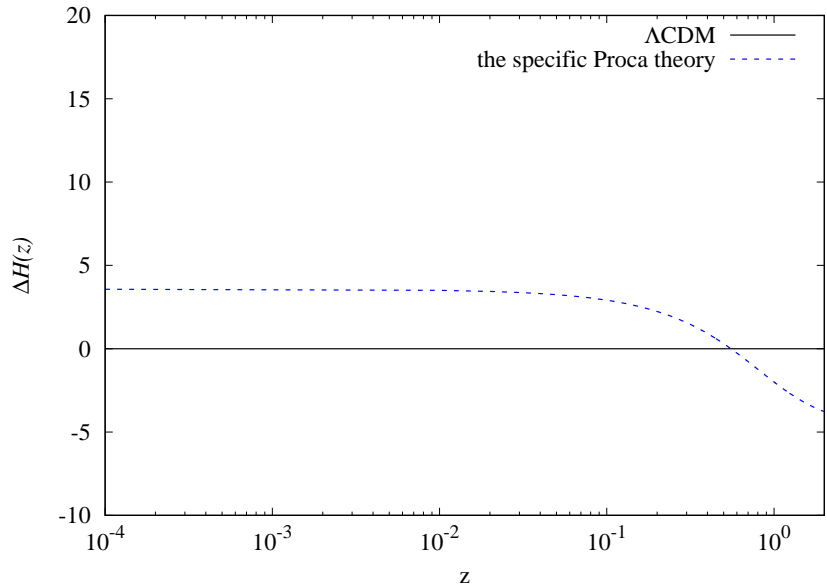}
	\caption{Evolution of Residues of Hubble parameter for the model and $\Lambda$CDM.}
	\label{fig:Hzresidue}
\end{figure}

Using the fitting results, we plot the evolutions of the Hubble parameter in our model and $\Lambda$CDM in Fig.~\ref{fig:Hzproca} with the initial conditions given by best-fit values listed in Table~\ref{tab:fit}. The residue of $H(z)_{Proca}-H(z)_{\Lambda CDM}$ are plotted in Fig.~\ref{fig:Hzresidue}. We note that $H(z)_{Proca} > H(z)_{\Lambda CDM}$ at the low redshift region when $z \lesssim 0.55$. However, $H(z)_{\Lambda CDM}$ will surpass $H(z)_{Proca}$ at the higher redshift one when $z > 0.55$.

\section{Conclusion}
We have studied a specific dark energy model based on the generalized Proca theories by only including up to the cubic order in the Lagrangian.
We have noticed that the dark energy evolution does not depend on  the values of $b_2$ and $b_3$ but $s$.
We have shown the phantom-like behavior  of $w_{DE}$ and evolutions of $\Omega_{m,r,DE}$  in the model.
By using the {\bf CosmoMC} and  {\bf CAMB} packages and fitting with the observational data of 
 the CMB data form {\it Planck} 2018 and the Hubble constant  of $H_0=74.03 \pm 1.42$ $\rm km\;s^{-1}Mpc^{-1}$ from HST,
 we have constrained the cosmological parameters.
In particular, we have obtained that $| \chi^2_{Proca,reduced}-1|=0.0960$ ($0.0959$) and
$|\chi^2_{LCDM,reduced}-1|=0.0964$ ($0.1016$) when fitting with CMB (CMB+HST).

As the generalized Proca theory is able to alleviate the $H_{0}$ tension at the background level~\cite{deFelice:2017paw}, 
our specific choice of the cubic order in terms of   the vector field self-interactions in the Lagrangian
also prefers a larger value of $H_{0}$,
determined to be $H_{0}^{Proca}=71.80^{+1.07}_{-0.72}$ ($72.48^{+0.72}_{-0.60}$) 
$\rm km\;s^{-1}Mpc^{-1}$ and $H_{0}^{\Lambda CDM}=66.75^{+1.52}_{-0.73}$ ($69.13\pm 0.57$) 
 $\rm km\;s^{-1}Mpc^{-1}$ at $68 \%$ C.L. with CMB (CMB+HST). 
 The increased value  of $H_0$ in our model matches the local measurements of HST, SH0ES, H0LiCOW and Megamaser, resulting in that the $H_{0}$ tension is reduced to $\sim 1 \sigma$ by comparing with the HST data. Furthermore, we have plotted
 the evolution of $H(z)$ with the initial conditions given by the best-fit  from the global fitting results, and found that $H(z)_{Proca}> H(z)_{\Lambda CDM}$ at the low redshift with $z \lesssim 0.55$.

\begin{acknowledgments}
This work is supported in part by MoST (Grant No. MoST-107-2119-M-007-013-MY3)
and
the National Key Research and Development Program of China (Grant No. 2020YFC2201501).
\end{acknowledgments}


\section{Appendix: Perturbations}
In this Appendix, we briefly list all the perturbed quantities in tensor, vector, and scalar perturbations and outline the perturbation equations for our specific dark energy model  from the Generalized Proca theory, as in Refs.~\cite{DeFelice:2016yws,DeFelice:2016uil,deFelice:2017paw}. We also adopt the method described in Ref.~\cite{DeFelice:2016yws}. Specifically,  we first expand ~\eqref{GPaction} up to the second order in perturbations, 
and then vary the second-order action
with respect to the perturbed quantities to arrive at the perturbation equations.
To perturb the gravity part of the action, we take 
the perturbing line element~\cite{Kodama:1985bj,Mukhanov:1990me},
\begin{align}
ds^2=-(1+2\a)dt^2+2(\pa_i \chi + V_i)dtdx^i+a^2(t)(\delta_{ij}+h_{ij})dx^i dx^j,
\end{align}
and the Proca vector field $A^{\m}$,
\begin{align}
A^0&=\f(t)+\d\f,\\
A^i&=\frac{1}{a^2(t)}\d^{ij}(\pa_j \chi_V+E_j),
\end{align}
where ($\a ,\chi$) and ($\d\f,\chi_V$) are the scalar perturbations of the metric and
$A^{\m}$, respectively, while $V_i$, $h_{ij}$, and $E_{i}$ satisfy the  conditions:
\begin{align}
\pa^{i}V_i&=0,\\
\pa^{i}h_{ij}&=0, \quad h^i{}_i=0,\label{TTgauge}\\
\pa^{i}E_i&=0.
\end{align}
In addition, we consider the perturbations of the matter
action by using the Schutz-Sorkin action~\cite{Schutz:1977df,Brown:1992kc},
\begin{align}
S_M=-\int d^4x [\sqrt{-g}\r_M(n)+J^{\m}(\pa_{\m}l+\mca_1\pa_{\m}\mcb_1+\mca_2\pa_{\m}\mcb_2)]
\end{align}
where $\r_{M}$ depends on the number density of the fluid $n$ defined by
\begin{align}
n=\sqrt{\frac{g_{\a\b}J^{\a}J^{\b}}{g}}.
\end{align}
We note that in this action, the pressure $P_{M}$ is related to $\r_{M}$ by~\cite{Schutz:1977df,Brown:1992kc}
\begin{align}
P_M=n_0\r_{M,n}-\r_M
\end{align}
with $n_0$ the number density of the fluid in the background. 
Besides, $l$ is a scalar, whereas $J^{\m}$ is the vector field of weight one,
$\mca_{1, 2}$ and  $\mcb_{1, 2}$ are scalars whose perturbations are meant to describe the vector modes.
For the FLRW background, $J^0$ corresponds to the total fluid number
$\mathcal{N}_0$.
They can be expressed as~\cite{DeFelice:2016uil}
\begin{align}
l&=-\int^{t}\r_{M,n}dt'-\r_{M,n}v,\\
J^0&=\mathcal{N}_0+\d J,\\
J^i&=\frac{1}{a^2}\d^{ik}(\pa_k\d j+W_k),
\end{align}
where $W_k$ also obeys the transverse condition,
\begin{align}
\pa^k W_k=0.
\end{align}
For the quantities of $\mca_{1, 2}$ and  $\mcb_{1, 2}$, we choose
the simplest forms satisfying the required properties for the vector mode as in Ref.~\cite{DeFelice:2016uil}, given by
\begin{align}
\mca_1=\d \mca_1(t,z), \quad \mca_2=\d \mca_2(t,z), \quad
\mcb_1=x+\d \mcb_1, \quad \mcb_2=y+\d \mcb_2(t,z),
\end{align}
where $\d \mca_i$ and $\d \mcb_i$ are perturbed quantities.

\subsection{Tensor perturbations}
For the tensor perturbation $h_{ij}$ to the metric, we use the transverse and traceless conditions \eqref{TTgauge}. To be specific, we express the components of $h_{ij}$ to be $h_{ij}=h_+ e_{ij}^{+}+h_{\times} e_{ij}^{\times}$, where $e_{ij}^{+}$ and $e_{ij}^{\times}$ obey the relations 
of $e_{ij}^{+}(\bs{k})e_{ij}^{+}(\bs{-k})^{\ast}=1$, $e_{ij}^{\times}(\bs{k})e_{ij}^{\times}(\bs{-k})^{\ast}=1$, and $ e_{ij}^{+}(\bs{k})e_{ij}^{\times}(\bs{-k})^{\ast}=0$ in the Fourier space with $\bs{k}$ being the comoving wave number. To obtain the tensor perturbation equations for our model, we first expand the action \eqref{GPaction} up to the second order ~\cite{DeFelice:2016yws,DeFelice:2016uil,deFelice:2017paw},
\begin{align}\label{ST2}
S^{(2)}_T=\sum_{\l=+,\times}\int dt \, d^{3}x \, M_{pl}^{2}\frac{a^3}{8}\Big[\dot{h}_{\l}^2-\frac{1}{a^2}(\pa h_{\l}^2)\Big].
\end{align}
Vary the above action~\eqref{ST2}, we find that 
\begin{align}
\ddot{h}_{\l}+3H\dot{h}_{\l}+\frac{k^2}{a^2}h_{\l}=0,
\end{align}
where $k=\left|\bs{k}\right|$.
We note that this is identical to the one in general relativity.

\subsection{Vector perturbations}
For the vector perturbations, the dynamical field $\ti{Z}_i$ can be expressed in terms of the combination of $E_i$ and $V_i$,
given by
\begin{align}
\ti{Z}_i=\frac{1}{a}[E_i+\f(t)V_i]\,.
\end{align}
Due to
the transverse condition, $\pa^i \ti{Z}_i=0$, we can choose that  $\ti{Z}_i=(\ti{Z}_1(t,z),\ti{Z}_2(t,z),0)$ without loss of generality. 
Similar to the case in the tensor perturbations, by expanding the action \eqref{GPaction} up to  the second order, taking the small-scale limit, and plugging in Eqs.~\eqref{G(X)} and \eqref{bpvalue}, the resulting action becomes~\cite{DeFelice:2016uil}
\begin{align}
S^{(2)}_V\approx\int dt \, d^{3}x \, \sum_{i=1}^{2} \frac{a^3}{2}\Big[\dot{\ti{Z}}_{i}^2+\frac{k^2}{a^2} \ti{Z}_{i}^2\Big],
\end{align}
which is similar to \eqref{ST2}.
As a result, the equations of motion for the vector perturbations are also similar to those in the tensor perturbations,
\begin{align}
\ddot{\ti{Z}}_i+3H\dot{\ti{Z}}_i+\frac{k^2}{a^2}\ti{Z}_i=0.
\end{align}

\subsection{Scalar perturbations}
For the scalar perturbations, the dynamical fields are $\psi=\chi_V+\f(t)\chi$ and $\d\r_M$. Expanding the action \eqref{GPaction} up to the second order in our choices of the parameters, one gets that~\cite{DeFelice:2016uil}
\begin{align}\label{SS2}
S^{(2)}_S&=\int dt d^3x \,a^3\bigg\{
-\frac{n_0 \r_{M,n}}{2a^2}(\pa v)^2
+\bigg[n_0 \r_{M,n}\frac{\pa^2\chi}{a^2}
-\dot{\delta}\r_M-3H(1+c_M^2)\delta \r_M\bigg]v\nonumber\\
-&\frac{c_M^2}{2n_0 \r_{M,n}}(\d\r_M)^2
-\a\d\r_M+2\frac{\f^2}{a^2}(\pa\a)^2
-(3M_{pl}^2H^2+b_2\f^2)\a^2\nonumber\\
-&\bigg[3b_2\f\d\f
+\frac{2\f}{a^2}\pa^2(\d\f)
+\frac{2\f}{a^2}\pa^2\dot{\psi}
+\frac{b_2 \f}{3a^2H}\pa^2 \psi
\bigg]\a\nonumber\\
-&\frac{(\pa\d\f)^2}{2a^2}-2b_2(\d\f)^2
-\bigg[\frac{b_2}{3H}\psi+\dot{\psi}\bigg]
\frac{\pa^2(\d\f)}{a^2}\nonumber\\
-&\frac{(\pa\dot{\psi})^2}{2a^2}
+\frac{b_2\dot{\f}}{6\f Ha^2}(\pa \psi)^2
+\bigg[(-2M_{pl}^2 H+\frac{b_2\f^2}{3H})\a
+\frac{b_2\f}{3H}\d\f\bigg]\frac{\pa^2\chi}{a^2}\bigg\}\,,
\end{align}
where $c_M$ corresponds to the matter propagation speed, given by~\cite{DeFelice:2016uil}
\begin{align}
c_M^2=\frac{n_0\r_{M,nn}}{\r_{M,n}},
\end{align}
and the matter perturbation $\d\r_M$ is defined by
\begin{align}
\d\r_M=\frac{\r_{M,n}}{a^3}\d J=\frac{\r_M+P_M}{n_0a^3}\d J.
\end{align}
Vary the action \eqref{SS2}, the equations of motion in the Fourier space for $\a, \chi, \d\f, v, \d\psi$, and $\d\r_M$ are given by
\begin{align}
&\d\r_M+(6M_{pl}^2H^2+2b_2\f^2)\a+3b_2\f\d\f+\frac{k^2}{a^2}\bigg[\mathcal{Y}-2M_{pl}^2H\chi+\frac{b_2\f}{3H}(\f\chi-\psi)\bigg]=0,\\
&(\r_M+P_M)v+\bigg(\frac{b_2 \f^2}{3H}-2M_{pl}^2 H\bigg)\a+\frac{b_2\f}{3H}\d\f=0,\\
&3b_2\f^2+4b_2\f\d\f+\frac{k^2}{a^2}\bigg[\frac{1}{2}\mathcal{Y}+\frac{b_2\f}{3H}(\f\chi-\psi)\bigg]=0,\\
&\dot{\d}\r_M+3H(1+c_M^2)\d\r_M+\frac{k^2}{a^2}(\r_M+P_M)(\chi+v)=0,\\
&\dot{\mathcal{Y}}+\bigg(H-\frac{\dot{\f}}{\f}\bigg)\mathcal{Y}+
\frac{2b_2}{3H}(\f^2\a+\dot{\f}\psi)+\frac{2b_2\f}{3H}\d\f=0,\\
&\dot{v}-3Hc_M^2v-c_M^2\frac{\d\r_M}{\r_M+P_M}-\a=0,
\end{align}
respectively, where 
\begin{align}
\mathcal{Y}\equiv-2\f\dot{\psi}-2\f\d\f-4\a\f^2.
\end{align}


\end{document}